\documentclass[conference]{IEEEtran}
\IEEEoverridecommandlockouts
\usepackage{cite}
\usepackage{algorithmic}
\usepackage{graphicx}
\usepackage{textcomp}
\usepackage{listings}
\def\BibTeX{{\rm B\kern-.05em{\sc i\kern-.025em b}\kern-.08em
    T\kern-.1667em\lower.7ex\hbox{E}\kern-.125emX}}

\lstdefinestyle{mystyle}{
    basicstyle=\ttfamily\footnotesize,
    breakatwhitespace=false,         
    breaklines=true,                 
    captionpos=b,                    
    keepspaces=true,                 
    numbers=left,                    
    numbersep=5pt,                  
    showspaces=false,                
    showstringspaces=false,
    showtabs=false,                  
    tabsize=2
}
\begin{document}

\title{NDVR: NDN Distance Vector Routing}

\author{\IEEEauthorblockN{Italo Valcy da Silva Brito}
\IEEEauthorblockA{
UFBA \\
italovalcy@ufba.br}
}

\maketitle

\begin{abstract}
Ad hoc mobile scenarios desire a lightweight routing protocol to propagate rapidly changing data reachability information in a highly dynamic environment. We are developing a distance-vector routing protocol that enables each node to selectively propagate a data reachability vector containing the named-data prefixes current reachable to their neighbors. In this report, we describe the implementation of NDVR (NDN Distance Vector Routing), discuss the rationale for the protocol design choices, and demonstrate a use case for the protocol to illustrate how the routing protocol can help NDN applications, specially in mobile adhoc scenarios.
\end{abstract}


\section{Introduction}

Ad hoc mobile scenarios desire a lightweight routing protocol to propagate rapidly changing data reachability information in a highly dynamic environment. The currently deployed Named-Data Networking (NDN) routing protocol, NLSR, is based on link-state algorithms, which require synchronization of the link-state database, which can be challenging to achieve in the above-intended scenario. We are developing a distance-vector routing protocol that enables each node to selectively propagate a data reachability vector containing the named-data prefixes current reachable to their neighbors.  Such reachability information can be propagated transitively, allowing all reachable nodes at the time to estimate their reachability to desired data  in a distributed and asynchronous manner.

This technical report presents our work in progress and prototype of the NDN Distance Vector Routing (NDVR) protocol\footnote{The code is available on GitHub: https://github.com/italovalcy/ndvr} \cite{brito2020}. The initial design of NDVR consists of the simplest possible way to propagate name prefix reachability information based on distance vector algorithms. The protocol enables dynamically identified neighbors to use NDN's Interest and Data packets to propagate routing updates and runs in two main phases: (i) dynamic neighbor discovery and (ii) distance-vector information exchange (dvinfo). NDVR prototype was developed on the ndnSIM simulation environment.

The report is organized as follows. In Section \ref{sec:design} we discuss
the rationale for the protocol design choices. In Section \ref{sec:evaluation} we demonstrate how the protocol can help a NDN application and present our preliminary evaluation. Finally, we conclude in Section \ref{sec:conclusion}.

\section{Design}
\label{sec:design}

NDVR was designed to be a lightweight routing protocol for wireless mobile adhoc mobile networks, which means that it propagates reachability information in a highly dynamic scenario, with minimal configuration and no additional synchronization requirements. NDVR's key design aspects are:
\begin{itemize}
\item Dynamically detect neighbors lightweight protocol to exchange reachability information (i.e., NDN name prefixes) using a distance-vector algorithm
\item  Provide support for highly dynamic mobile adhoc wireless network
\item Use the NDN trust model to secure the protocol data packets.
\end{itemize}

\begin{figure}[!ht]
	\centering
	\includegraphics[width=1\columnwidth]{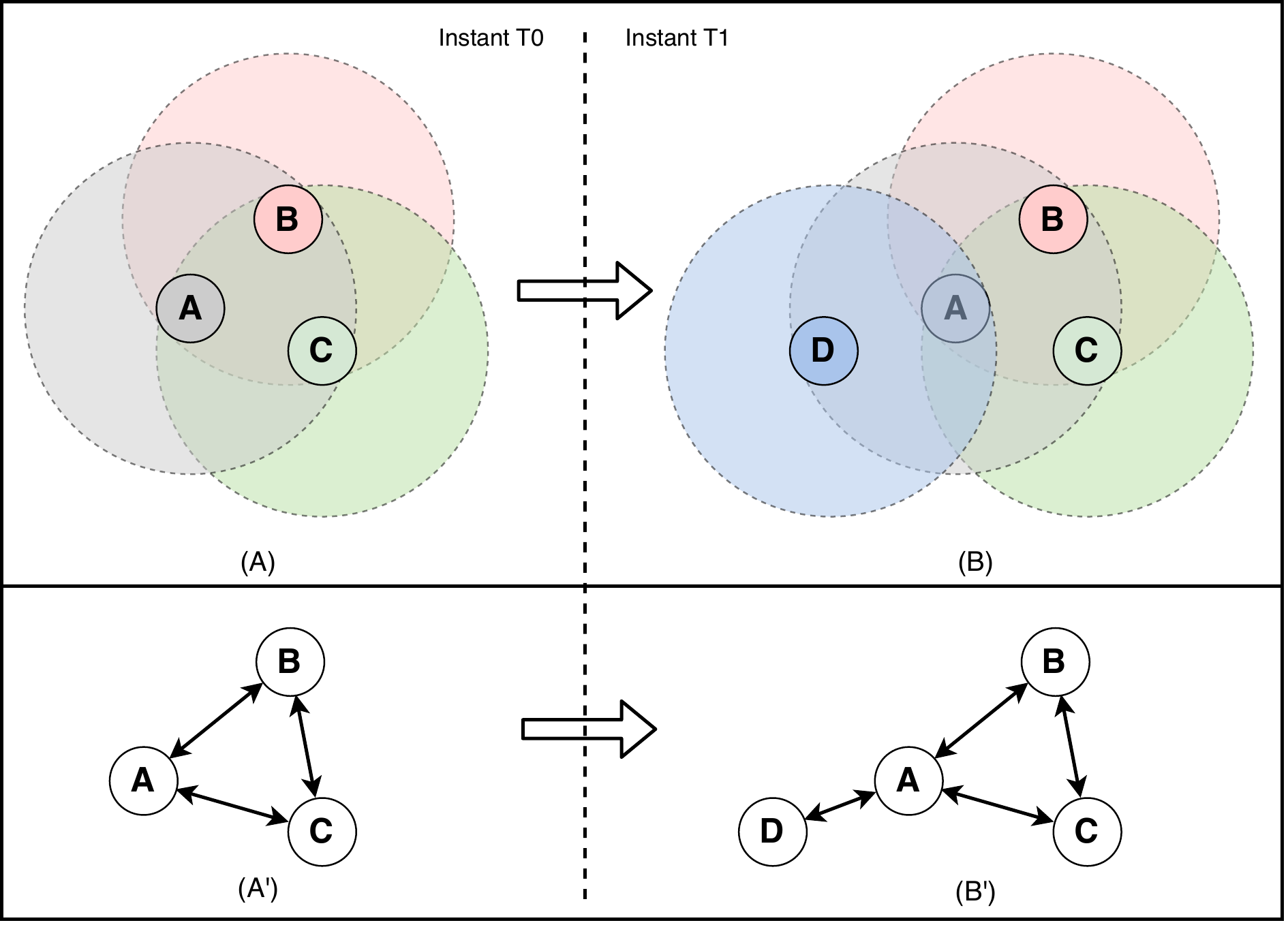}
	\caption{Mobile adhoc wireless network scenario where the NDVR protocol will be used.}
    \label{fig:topology-range}
\end{figure}

To better understand where and how the NDVR protocol can be used, consider the scenario illustrated in Figure \ref{fig:topology-range}. Figure \ref{fig:topology-range} represents an adhoc wireless scenario where, in the instant T0 (Figure \ref{fig:topology-range}A), there is a group of three nodes, A, B, and C. Each node has a wireless network interface configured in Adhoc mode. We assume a range wireless signal propagation model for simplicity, where the outer opaque circle represents the range of each node. For example, the gray filled outer circle is the wireless range for A, the pink filled outer circle is the wireless range for B, and the green filled outer circle represents the wireless range for C. Thus, in Figure 1A the nodes A, B, and C are interconnected (they are in the wireless range of each other). In the instant T1 (after T0), a mobile node D moves towards the group and reaches the wireless range of A, establishing a wireless link between D and A. It is worth notice that D does not establish a wireless association with either A or C because the transmission power of D is not strong enough to reach A's or C's reception antenna (i.e., A and C are not inside the blue circle representing D's transmission range) and vice-versa (i.e., D is not inside the green circle representing C's transmission range and D is not inside the pink circle representing A's transmission range). Figures \ref{fig:topology-range}A' and \ref{fig:topology-range}B' represent the same scenario using arrows to represent the wireless association between nodes. Thus, the arrow between A and B means that A is at the B's wireless range and B is at A's wireless range.

Each node in the scenario can participate in the routing process by (i) advertising local reachability information (e.g., name prefixes of NDN producer applications or repos running on the same node) or by (ii) learning and disseminating reachability information from other reachable neighbors. To illustrate the routing process, let's assume that every node advertises its local prefixes, i.e., node A advertises $/ndn/a$, node B advertises $/ndn/b$ so on.

We also assume that when the scenario first starts, the nodes don't know about each other; they don't have any information about the neighborhood. This assumption also applies to node mobility: when a node moves around the topology, we assume that the neighborhood is unknown and needs to be discovered. For example, when the topology starts in Figure 1A', nodes A, B, and C don't actually know they are neighbors. Indeed, they send periodic messages to announce themselves, and so they get to know the neighbors. Whenever a neighbor is detected, the node sends an Interest to retrieve the neighbor distance vector table with routing updates, if any.

\subsection{Naming schema}
\label{subsec:naming-schema}

A critical aspect of the design of NDVR is the naming schema and its relationship with the trust model. We took the hierarchical naming schema from NLSR and simplified it as much as possible to have the minimal naming components for the protocol features and enabling the proper data validation.

The protocol is designed so that each router is named hierarchically under the network name prefix where the router resides, followed by the actual router name. The router name is formatted in three parts: the network or organization name (e.g. $/google$, $/ufba$, $/ndn$), router command marker (i.e., $\%C1.Router$) and a router label (e.g., RouterA). For example, considering Figure \ref{fig:topology-range}B' and supposing all the wireless routers are running at the UFBA University campus, they would be named as $/ufba/\%C1.Router/A$, $/ufba/\%C1.Router/B$, $/ufba/\%C1.Router/C$ and $/ufba/\%C1.Router/D$. The router name is supposed to be unique on the network. It may be chosen from the hostname, highest ethernet MAC address from all wireless interfaces, or specially created (e.g., the name can be formatted to represent a hierarchical topology). The uniqueness of the router name is important for the trust model because it must match the name of the public key signing the protocol data packets (more information on Section \ref{sec:security}).

NDVR's naming design is composed of two main message types: EHLO and DVINFO. The EHLO message (Extended Hello) is used for neighbor discovery and to announce when the router has new updates on its distance vector information. The DVINFO message (Distance Vector Information) is used to exchange the distance vector table with the neighbors. Both messages are only exchanged in the neighborhood, using the following naming schema:
\begin{itemize}
    \item $/localhop/ndvr/ehlo/<RouterName>/<\#prefixes>/<\#ver>/<digest>$: this name prefix is used to exchange EHLO messages between neighbors. Each router sends periodic interest packets through all its faces. The ehlo interval can be configured and defaults to one second to better deal with highly dynamic MANETs. To receive the neighbor's ehlo messages, each router has a pre-installed FIB entry that forwards the $/localhop/ndvr/ehlo$ interests to the NDVR application. The $/localhop$ defines the scope in which the interest will be forward: only in the neighborhood \cite{ndnwiki}. The name component $\#prefixes$ is an integer to represent the number of name prefixes the router has to advertise. The number of prefixes is used to decide whether the node should fetch or not the neighbor's distance-vector ($\#prefixes$ > 0). Furthermore, if a router - let's say router A - has more prefixes to advertise than its neighbor - let's say router D -, the neighbor should request the distance-vector first. This is a prioritization strategy we adopt to avoid redundant distance-vector exchange within a group of nodes. The name components $\#ver$ and $digest$ represent the distance vector table state, and they are used to help the neighbors fetch and synchronize their distance vector information.
    \item $/localhop/ndvr/dvinfo/<RouterName>/<\#ver>$: this namespace is used to exchange on-demand distance vector information (DVINFO) between neighbors. As soon as a node, let's say A, detects that a neighbor, let's say B, has new reachability information (i.e., $\#prefixes$ is greater than zero and the $\#ver$ is newer than any previous - if any - and the digest is different from the local distance vector table), then A sends an Interest for B's distance vector (e.g., $interest.name = /localhop/ndvr/dvinfo/ufba/\%C1.Router/B/1$) and B replies with its distance vector table as a signed data packet. Once A receives the data packet containing B's DVINFO, A starts the validation process. The validation process may require B's key if not previously seen. Thus, A will also send an interest for B's key according to the Key Locator from the DVINFO packet. Once A finishes the validation process successfully, A processes B's DVINFO using the well-known Bellman-Ford Distributed algorithm with sequence numbers \cite{perkins1994}.
    \item $/<network>/<RouterName>/KEY$: this namespace is used to fetch the keys that sign DVINFO data packages. Following the same approach as NLSR, the nodes obtain the neighbors key by using a direct fetch mechanism.
\end{itemize}

\subsection{Neighbor detection and routing updates propagation}

\subsubsection{Neighbor detection}

\begin{figure}[!ht]
	\centering
	\includegraphics[width=1\columnwidth]{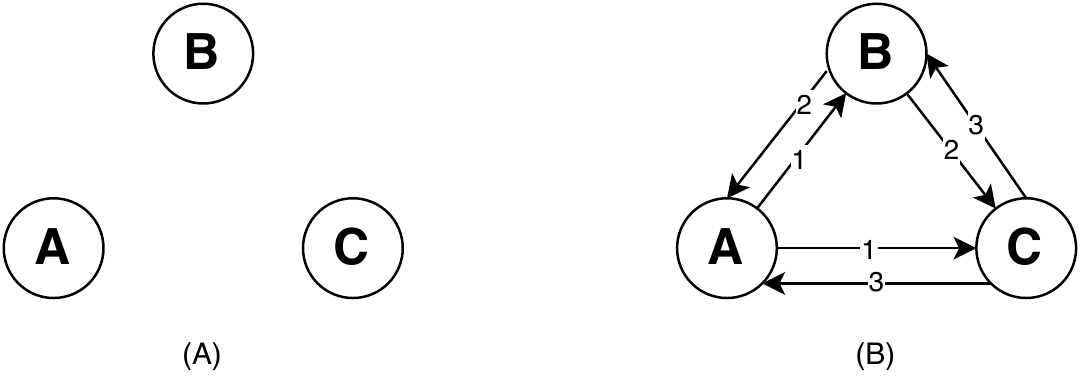}
	\caption{Scenario to illustrate the neighbor discovery process.}
	\label{fig:neighbor-discovery}
\end{figure}

The neighbor detection process is based on each node periodically sending Interest packets on its non-local faces. All nodes are pre-configured with a FIB entry to forward interests matching the namespace $/localhop/ndvr/ehlo$ to the NDVR application (the pre-configuration is done by creating a FIB entry on the NDVR application startup routine). Figure \ref{fig:neighbor-discovery} illustrates the neighbor detection since the moment zero when the network just started. The nodes don't know about each other's existence (Fig. \ref{fig:neighbor-discovery}A) until they all send periodic EHLO interests (Fig. \ref{fig:neighbor-discovery}B). 

Note that even though we enumerate the messages sent by A as 1, by B as 2, and by C as 3, there is no specific order in which those events may occur. Each node runs its instance of the NDVR protocol application, and they send the periodic EHLO by themselves. It is worthy of mentioning, though, that depending on the host communication model (i.e., the transport method - ethernet broadcast, unicast, udp, udp+multicast, etc.), if all nodes send their interest at the same time through the wireless shared medium, the network may be impacted by collisions and packet loss (especially on the ndnSIM simulation environment, but also in real environments depending on the communication model). A simple and effective strategy is to randomly wait a couple of milliseconds before starting the periodic EHLO routine. The periodic EHLO messages are based on the $ehloInterval$ configuration parameter, which defaults to one second.

Each node keeps a list of neighbors, along with their most recent DVINFO version, the last time the neighbor was seen, the first time has seen, and the face id from which the neighbor is accessible (for multiple face scenarios).

When a node moves around the topology, the NDVR routing protocol must quickly detect new neighbors and neighbors that moved away, then update its distance vector table. The periodic EHLO message works just as it is for detecting new neighbors. On the other hand, to remove an adjacency, the NDVR protocol waits a certain time without receiving EHLO messages from a node to consider the node timed out and no longer neighbor. The timeout for periodic EHLO messages is based on the $ehloInterval$ and configured through the parameter $ehloMultiplier$, which defaults to 2, meaning the timeout will be twice the ehlo interval value. Configuring the $ehloMultiplier$ to 1 leads to frequent neighbor removal due to variations on the wireless medium propagation delay, and higher values may postpone the update of association status for mobile nodes.

\subsubsection{Routing updates propagation}

Routing updates propagation is the process in which the nodes notify their neighbors about new name prefixes on-demand, and the neighbors fetch distance vector information. The on-demand routing updates propagation means that a node sends routing announcements to its neighbors only when/if the node produces new data and publishes reachability information about the produced data through the NDVR routing protocol. For example, in the previous scenario from Figure 1, node A runs the NDVR protocol application and an NDN producer application, which produces data for the name prefix /ndn/a. The same applies for node b, with the name prefix /ndn/b, nodes c and d, with /ndn/c and /ndn/d, respectively. When the NDN producer application produces new data, it notifies the local NDVR process to announce the newly produced data. Then, the NDVR protocol application notifies its neighbors about new reachability information by sending the next EHLO with the number of prefixes, an incremented version, and a new digest. From the latest version and digest, the neighbors send interest for the DVINFO.

When an NDVR node sends an interest to notify new reachability information, the wireless shared medium - especially in adhoc mode with broadcast faces - multiple nodes in the same communication range may receive the same interest notification and request the DVINFO at the same time. NDVR adopts a prioritization approach where only a subgroup of the neighbors sends interest for the DVINFO right away. Other nodes in the communication range wait a backoff time\footnote{Backoff time here refers to the strategy of decreasing the rate of some process to gradually find an acceptable rate. In particular, it was used to wait for some period in order to avoid collisions or redundant interest.} before sending the interest, which will hopefully be satisfied from their opportunistic cache populated from the answers for the priority subgroup of neighbors. The size of the priority subgroup of neighbors defaults to two (primary and secondary) to guarantee at least one request even under packet loss environments. The subgroup of neighbors is chosen using a rounding robin approach so that no single node is overloaded to request the DVINFO every time.

\begin{figure}[!ht]
	\centering
	\includegraphics[width=1\columnwidth]{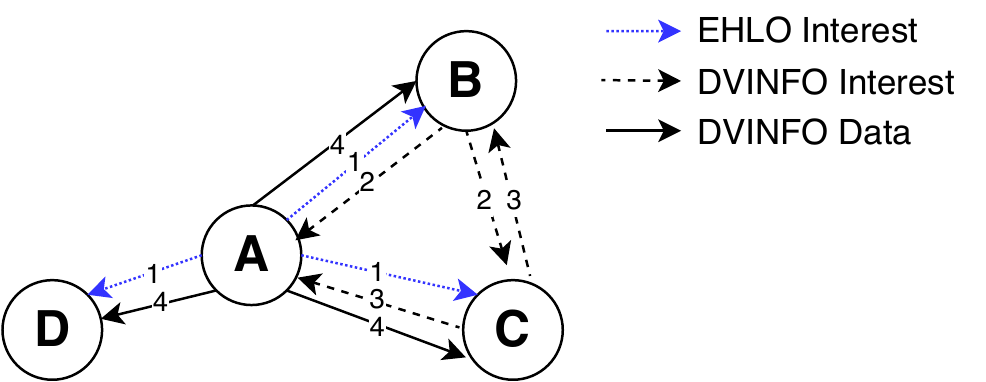}
	\caption{DVINFO exchange between nodes. A send a notify interest, B and C request A's DVINFO and B, C, D receive the reply (D overhears)
.}
    \label{fig:dvinfo-exchange}
\end{figure}

Figure \ref{fig:dvinfo-exchange} illustrates the above approach in the same scenario presented before (with all four nodes, A, B, C, and D). In this example, node A has new reachability information, and as so A sends an EHLO (1) with an increased version. In the EHLO interest sent by A, it chooses B and C to be the requesters of its DVINFO; those are A's priority subgroup. All other neighbors not listed in A's priority subgroup - in this case, D - will wait a time before sending its interest. Since B and C are in A's priority subgroup, they will send an interest right away to fetch A's DVINFO content (2 and 3). Router A will receive two interests from B and C (unless there is packet loss); however, the order and inter-arrival time is unknown - it will depend on the wireless channel. Note that C also receives the DVINFO interest from B (2), and B also gets the DVINFO interest from C (3); however, since they are not producers for that name prefix and due to the localhop scope, they drop the interest. To avoid answering the same interest twice (one from the application and another from the cache), A also delays its content for some milliseconds increasing the change to avoid extra overhead. When A sends the data packet with its DVINFO table (4), all nodes in the communication range will receive it. Thus, D will overhear A's DVINFO and opportunistically save this data on its content store. Later, when D finishes the wait time, it will fetch A's DVINFO from the content store. To enable a node to overhear NDVR DVINFO data packets, we deployed a custom "unsolicited data policy" which caches NDVR data packets (the default policy is to drop unsolicited data packets).

Table \ref{table:dvinfo-pkt-flow} illustrates the messages exchanged in the scenario presented in Figure \ref{fig:dvinfo-exchange}. From the column Name, which represents a simplified name in the NDN packet, you can see how A represents its priority subgroup - by using application parameters. We can also see how B and C generate two interests for the same name prefix (different nonces). 

\begin{table}[ht]
\begin{tabular*}{\columnwidth}{p{0.15cm}|p{0.7cm}|p{0.6cm}|p{4cm}}
\# & Type     & Sender & Name (simplified example)  \\
\hline
1  & Interest & A & /localhop/ndvr/ehlo/ufba/\%C1.Router/A/1/10/d1/ params=/ufba/\%C1.Router/B\&/ufba/\%C1.Router/C \\
\hline
2  & Interest & B      & /localhop/ndvr/dvinfo/ufba/\%C1.Router/A/10? Nonce=100  \\
\hline
3  & Interest & C      & /localhop/ndvr/dvinfo/ufba/\%C1.Router/A/10? Nonce=200  \\
\hline
4  & Data     & A      & /localhop/ndvr/dvinfo/ufba/\%C1.Router/A/10                                         
\end{tabular*}
\caption{Packet flow representing the DVINFO exchange process from Figure \ref{fig:dvinfo-exchange}}
\label{table:dvinfo-pkt-flow}
\end{table}

To encapsulate the distance vector table into the NDN data content, NDVR serializes the distance vector information using Google's protocol buffer (protobuf) standard\footnote{https://developers.google.com/protocol-buffers/docs/cpptutorial}. Among many other serialization strategies, protobuf showed to be very efficient and flexible. The information encapsulated into the DVINFO data packet is: the reachable name prefix cost to reach the name prefix and sequence number. Like NLSR, the reachable name prefix refers to name prefixes that a router or its directly connected neighbors can reach. In other words, they either produce or host content with names that fall under the advertised name prefixes. The cost to reaching the name prefix is a metric provided by NDVR to qualify the path towards the name prefix's origin. The cost can be based on the hop count or any other strategy (e.g., cumulative delay, residual bandwidth, signal strength, etc.). NDVR's cost defaults to hop count. Finally, the sequence number is used to represent the freshness of the route and avoid counting-to-infinity problems \cite{mohapatra2004} (also referred to as short-term and long-term routing loops, which are well-known problems present on the Distributed Bellman-Ford (DBF) basic distance vector algorithm). The usage of sequence numbers on mobile adhoc networks is well discussed in the literature [4], and one of its primary examples is the DSDV routing algorithm \cite{perkins1994}. The basic idea of using sequence numbers is to tag each name prefix on the origin router with a monotonically increasing number and use this number later in the routing propagation process (as part of the DBF algorithm) to determine the relative freshness of reachability information (the highest sequence number is preferred). We refer to \cite[Section 3.3.1]{mohapatra2004} and \cite{perkins1994} for more details on the sequence numbers for MANET routing algorithms.

The algorithm for processing the distance vector information is shown in Listing \ref{lst:ndvr-algorithm}. It is mostly the basic DBF algorithm with sequence numbers. The main difference is the incoming NDN face as the next hop for later installing the resulting routes on the FIB. The function calculateCost() increments the hop count as the routing selection metric.

\begin{lstlisting}[caption=Simplified algorithm for processing the DVINFO (mostly based in DBF algorithm with sequence number)., label={lst:ndvr-algorithm}]
Node i receives a DvInfo message from neighbor j
For each name prefix d on DvInfo message:
	cost_i_d = calculateCost(j, cost_j_d)
	if isNewPrefix(d) or (seqNum_i_d < seqNum_j_d) or ((seqNum_i_d == seqNum_j_d) and (cost_i_d > cost_j_d)) then
		learnPrefix(d, nextHop = j, cost = cost_i_d, seqNum = seqNum_j_d)
\end{lstlisting}

\subsection{Security}
\label{sec:security}

Every NDN Data packet is digitally signed, covering the name, content, and metadata. The public key used to sign the NDN Data packet is specified in the actual packet, in the field KeyLocator. The data packet validation follows a workflow where i) the client fetches the key specified on the KeyLocator, ii) the client validates the public key properties (expiration, revocation, cryptography algorithms) and authenticates the key against the trust model (trust anchors and validation rules), iii) the client verifies the signature on the data packet. NDVR leverages data validation for reachability exchange by deploying a simplified trust model and validation rules.

\begin{figure}[!ht]
	\centering
	\includegraphics[width=1\columnwidth]{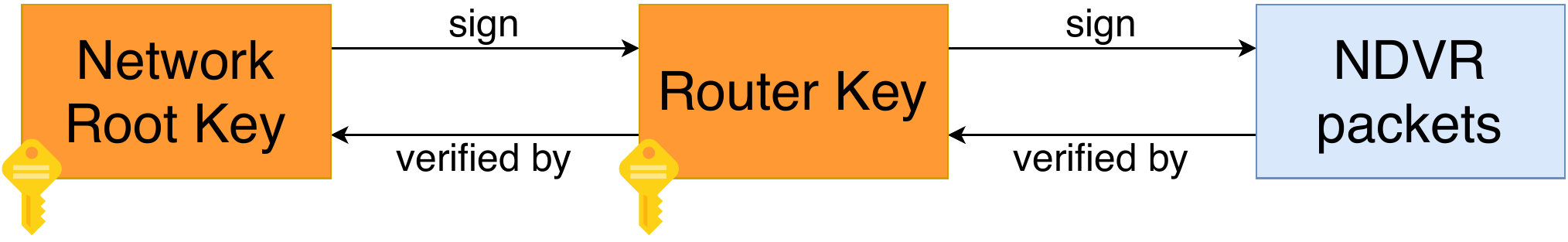}
	\caption{Illustration of the relationship between the keys in NDVR trust model.}
    \label{fig:ndvr-trust-model-rel}
\end{figure}

NDVR uses a two-level hierarchical trust model to represent the intra-domain routing structure, as shown in Table \ref{table:ndvr-trust-model}. There are only two levels on the trust chain: the Network Key, which is our trust anchor, and as so it is supposed to be pre-installed on all nodes; the Router Key, which is used to sign the data packets directly. The Network Key is used to validate the Router Key, which turns to be used to validate NDVR data packages (i.e., DVINFO). This relationship between all those keys and the trust model is shown in Figure \ref{fig:ndvr-trust-model-rel}. 

\begin{table}[ht]
\begin{tabular*}{\columnwidth}{p{3cm}|p{3cm}}
Key owner & Key name  \\
\hline
Network (Trust Anchor - pre-installed) & /<network>/KEY \\
\hline
Router  & /<network>/<routerName>/KEY \\
\end{tabular*}
\caption{Simplified Trust Model for NDVR.}
\label{table:ndvr-trust-model}
\end{table}

Listing \ref{lst:validation} shows the validation rules applied to NDVR packets. There are three validation rules, starting from bottom to top: i) in lines 57-61 is the validation rule for the trust anchor, which trust is assumed (not derived); ii) in lines 29-55, validation rule to ensure the router key is signed by the network key (trust is derived); and finally, iii) lines 1-27 which include the validation rule for the DVINFO data packets, which must be signed by a router key with the same name of the DVINFO routerName component name (trust also derived).

\begin{lstlisting}[caption=Validation rules for NDVR signed packets, label={lst:validation}]
rule
{
  id "DvInfo data should be signed by Router's key"
  for data
  filter
  {
    type name
    regex ^<localhop><ndvr><dvinfo><><%C1.Router><><><><>$
  }
  checker
  {
    type customized
    sig-type rsa-sha256
    key-locator
    {
      type name
      hyper-relation
      {
        k-regex ^([^<KEY>]*)<KEY><>$
        k-expand \\1
        h-relation equal
        p-regex ^<localhop><ndvr><dvinfo>(<><%C1.Router><>)<><><>$
        p-expand \\1
      }
    }
  }
}

rule
{
  id "Router's key should be signed by Network's key"
  for data
  filter
  {
    type name
    regex ^<><%C1.Router><><KEY><><><>$
  }
  checker
  {
    type customized
    sig-type rsa-sha256
    key-locator
    {
      type name
      hyper-relation
      {
        k-regex ^([^<KEY>])<KEY><>$
        k-expand \\1
        h-relation equal
        p-regex ^(<>)<%C1.Router><><KEY><><><>$
        p-expand \\1
      }
    }
  }
}

trust-anchor
{
  type file
  file-name "./trust.cert"
}
\end{lstlisting}

\section{Preliminary Evaluation}
\label{sec:evaluation}

The preliminary evaluation was based on two experiments to validate how the routing protocol can help different aspects of a NDN scenario: using the routing protocol on a Data Synchronization scenario and using the routing protocol to help the forwarding strategy.

\subsection{Routing and Data Synchronization}

To evaluate how routing can help the data synchronization function, this experiment compares NDVR routing protocol and a simple producer/consumer NDN application versus DDSN (Distributed Dataset Synchronization over disruptive Networks) \cite{li2019distributed}. DDSN was chosen to be compared because, even though it is a synchronization strategy (not routing), it does provide some features for mobile adhoc networks that a routing strategy can also provide. To provide a fair comparison, NDVR routers also run a simple NDN sync application that produces and fetch data upon receiving a new name prefix from NDVR's routing announcements. Thus, the sync protocol can be seen as i) state synchronization (i.e., learn the latest data generated by others) is provided by NDVR by propagating reachability information about each name prefix of latest generated data; ii) data synchronization (i.e., fetch the newly learned data) is provided by the simple sync NDN application. We are working on future evaluations more suitable for the routing strategy specifically. 

The evaluation scenario is similar to the proposed on DDSN paper \cite{li2019distributed}: 20 mobile nodes on an 800m x 800m topology with mobility based on the Random Walk Mobility Model, speed ranges from 1m/s to 20m/s, direction ranges from $0$ to $2\pi$, each node moves along the same path for 20s before changing its direction and speed, network device IEEE 802.11b 2.4GHz (transmission rate of 11Mbps), Wi-Fi propagation model based on a maximum range of 60m. Each node generates data following a Poisson distribution with $\lambda = 40s$ on average, and its data generation process lasts for 800s.

\begin{figure}[!ht]
	\centering
	\includegraphics[width=1\columnwidth]{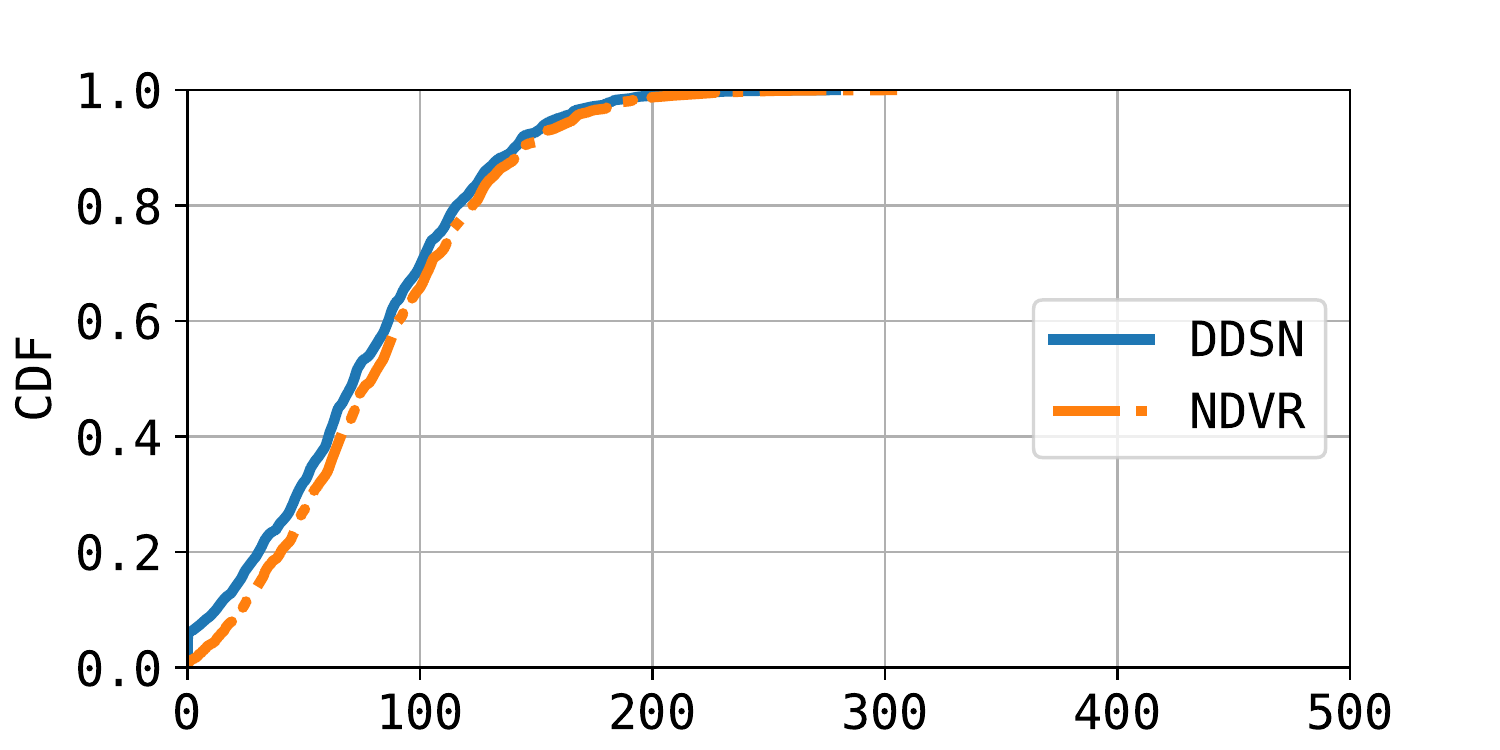}
	\caption{Data retrieval delay, NDVR+SimpleSync and DDSN.}
    \label{fig:data-retrieval-delay}
\end{figure}

\begin{figure}[!ht]
	\centering
	\includegraphics[width=1\columnwidth]{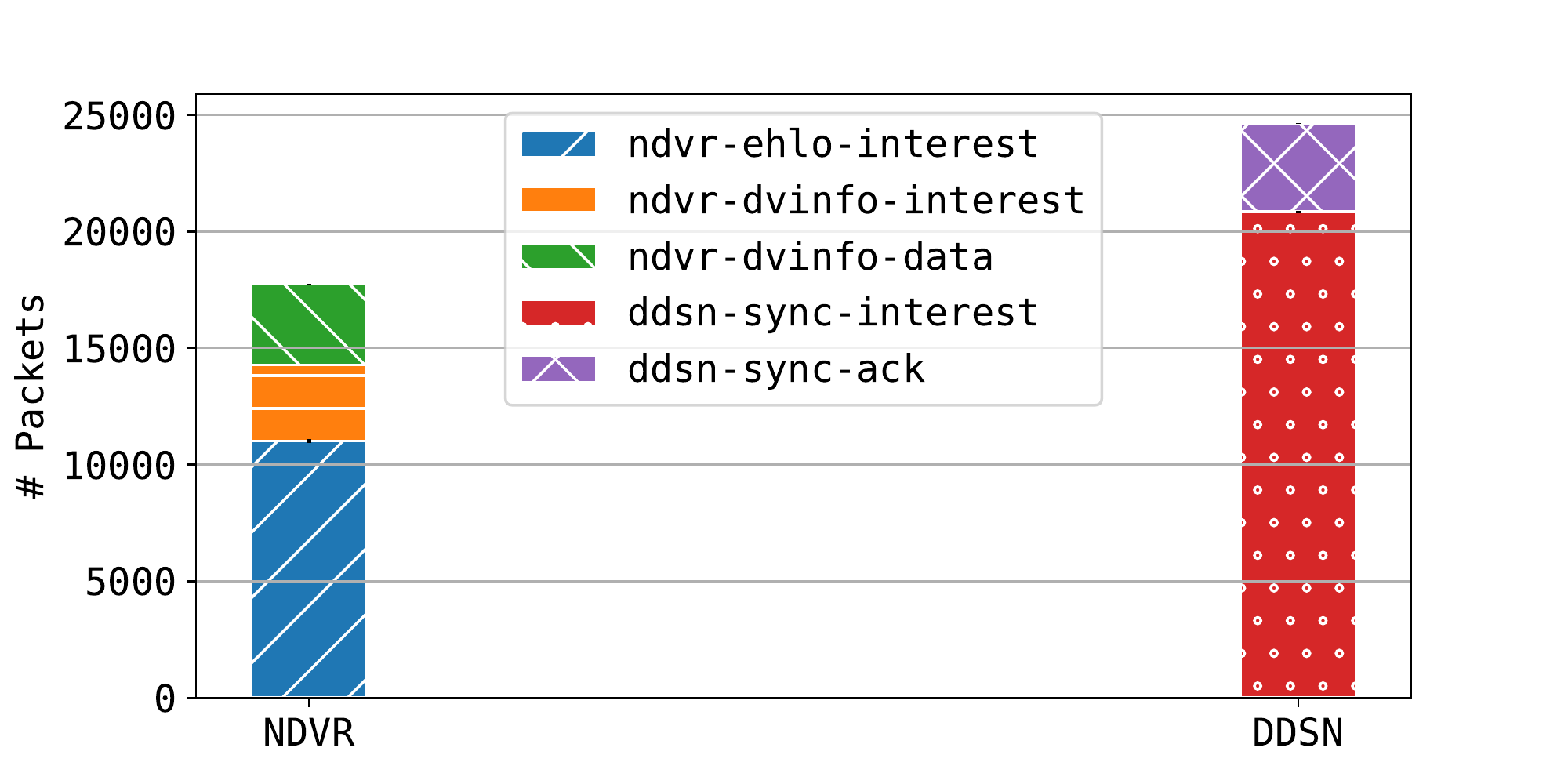}
	\caption{Protocol overhead in total number of packets, NDVR and DDSN.}
    \label{fig:overhead}
\end{figure}

Figure \ref{fig:data-retrieval-delay} presents the CDF graph of the data retrieval delay for DDSN and NDVR+SimpleProdCons (NDVR). And Figure \ref{fig:overhead} illustrates the protocol overhead (considering a 95\% confidence interval) in terms of the total number of packets. Figure 7 shows that NDVR and DDSN have very similar behavior considering the data retrieval delay, which accounts for both sync functions: state sync or reachability propagation; and data sync (data fetch). On the other hand, Figure 8 shows that DDSN has a higher protocol overhead considering the number of packets, which indicates that NDVR can save resources of the mobile nodes and also reduce the congestion/collisions in the wireless channel. In summary, despite being a well robust protocol, DDSN, and NDVR shows similar performance but with reduced overhead for NDVR.

\subsection{Routing and Forwarding strategies}

This experiment evaluates how the routing protocol can help a forward strategy in better doing its task of selecting a subset of next-hop and forward interests. We investigate four different configurations in the same scenario: i) multicast forwarding strategy with a default route\footnote{The default route means one FIB entry for the name $/$ and using the wireless adhoc face as next-hop}, ii) m-ASF (Adaptive Smoothed RTT-based Forwarding - ASF - for mobile networks \cite{masf2020}) and default route, iii) NDVR routing plus multicast forwarding, and iv) NDVR plus m-ASF. The scenario was based on a topology with 15 mobile nodes on a 300 m x 300 m area with a group mobility model based on Reference Point Group Mobility, the average group size of 3 nodes with a standard deviation $0.2$ and zero group change probability. Speed ranges from 1m/s to 20m/s, network device IEEE 802.11b 2.4GHz (transmission rate of 11Mbps), Wi-Fi propagation model based on a maximum range of 60m. Every node produces data on a different name prefix ($/ndn/dataSync/X$, X being the node id). They also consumes data from all other nodes, following a Constante Bit Rate traffic model, with $IDT = 100ms$ (Inter Departure Time) and $PS = 300 bytes$ (Payload size) during 100 seconds.

\begin{figure}[!ht]
	\centering
	\includegraphics[width=1\columnwidth]{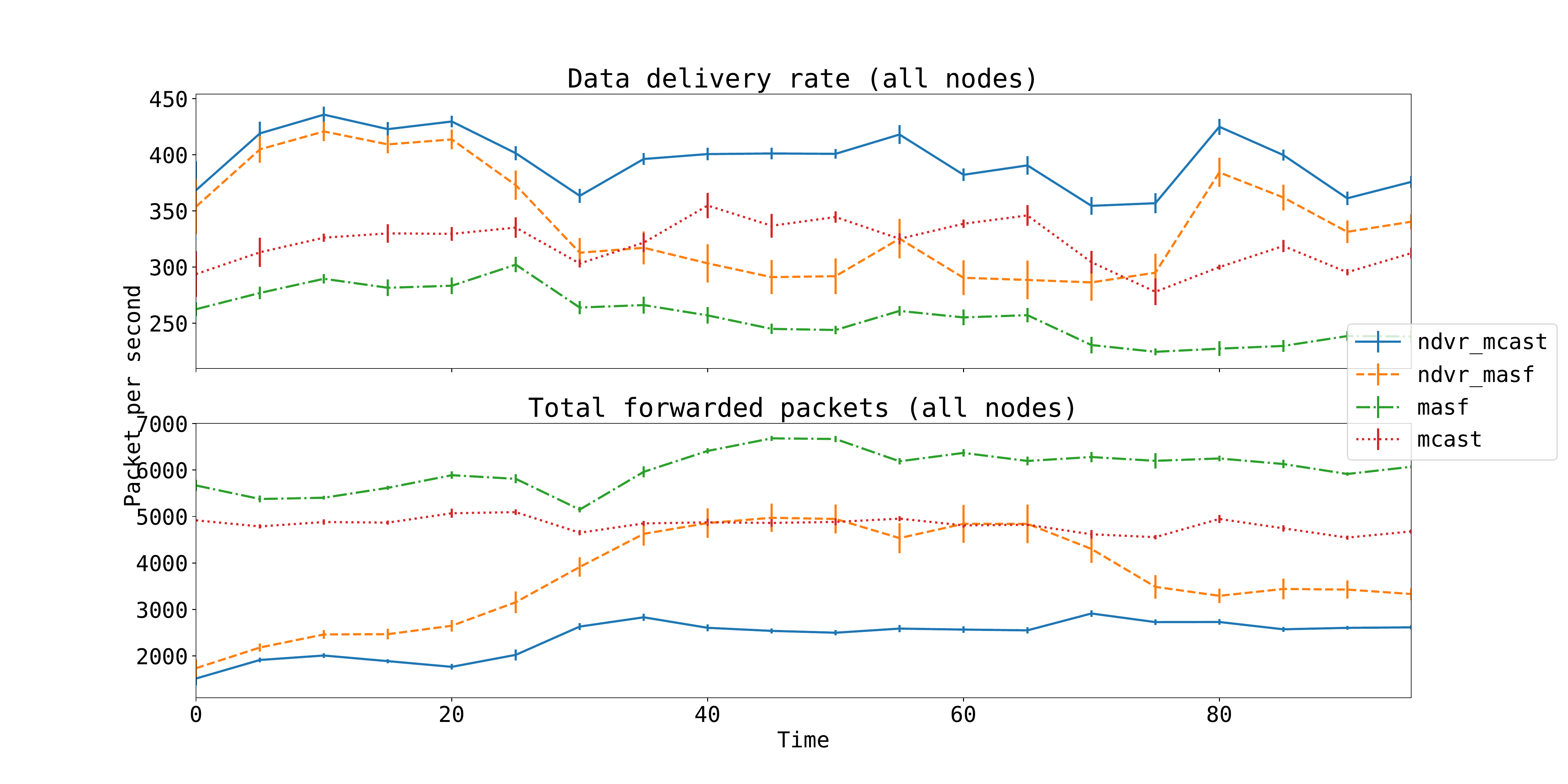}
	\caption{Data delivery rate and total forwarded packets among different strategies with and without NDVR routing protocol}
    \label{fig:routing-forwarding-data-total}
\end{figure}

Figure \ref{fig:routing-forwarding-data-total} presents the Data Delivery Rate and the Total Overhead (Total number of forwarded packets), both aggregated for all nodes. Regarding the data delivery rate, it is possible to see that both multicast and m-ASF forwarding strategies benefit from using the NDVR routing. The higher data delivery rate results from an efficient name reachability information propagation that saves the nodes from forwarding unnecessary interests when the data is not reachable. The total number of forwarded packets also reflects how the routing protocol reduces overhead and, consequently, reduces collisions/contention on the wireless channel. On average, for the data delivery rate, the multicast strategy improved from $320.39 \pm 9.42 pps$ to $395.18 \pm 11.90 pps$, and the m-ASF strategy improved from $256.75 \pm 10.38 pps$ to $339.78 \pm 22.07 pps$. For the overhead, the multicast reduced from $4820.57 \pm 71.09 pps$ to $2405.07 \pm 185.15 pps$ and the m-ASF reduced from $6010.57 \pm 194.07 pps$ to $3674.06 \pm 482.16 pps$.

\section{Conclusion}
\label{sec:conclusion}

This technical report presented the development of NDVR (NDN Distance Vector Routing), a routing protocol for NDN mobile adhoc networks. NDVR design consists in the simplest possible way to propagate name prefix reachability information based on distance vector algorithms. Neighbors are detected dynamically, and routing updates are propagated on-demand. Furthermore, NDVR from the wireless shared medium and NDN opportunistic caching characteristics to reduce the protocol overhead by using specific strategies when broadcasting protocol messages in a group of nodes. The simplified trust model and strict validation rules ensure the reachability information exchange is secure, allowing only trusted routers to participate in the routing propagation process.

NDVR was preliminarily evaluated using a prototype developed in the ndnSIM simulator and a comparison with the DDSN synchronization strategy and evaluated on the integration with multicast and m-ASF forwarding strategies. The results are promising, showing a comparable performance for data retrieval delay, efficient data reachability information propagation, and enhanced protocol overhead.

\section{Acknowledgments}
\label{sec:acknowledgments}

I want to thank my shepherd, professor Leobino Sampaio, for his valuable comments, suggestions, and thoughtful discussions. I also thank professor Lixia Zhang for the collaborations.

\bibliographystyle{IEEEtran}
\bibliography{main}

\begin{thebibliography}{1}
\providecommand{\url}[1]{#1}
\csname url@samestyle\endcsname
\providecommand{\newblock}{\relax}
\providecommand{\bibinfo}[2]{#2}
\providecommand{\BIBentrySTDinterwordspacing}{\spaceskip=0pt\relax}
\providecommand{\BIBentryALTinterwordstretchfactor}{4}
\providecommand{\BIBentryALTinterwordspacing}{\spaceskip=\fontdimen2\font plus
\BIBentryALTinterwordstretchfactor\fontdimen3\font minus
  \fontdimen4\font\relax}
\providecommand{\BIBforeignlanguage}[2]{{%
\expandafter\ifx\csname l@#1\endcsname\relax
\typeout{** WARNING: IEEEtran.bst: No hyphenation pattern has been}%
\typeout{** loaded for the language `#1'. Using the pattern for}%
\typeout{** the default language instead.}%
\else
\language=\csname l@#1\endcsname
\fi
#2}}
\providecommand{\BIBdecl}{\relax}
\BIBdecl

\bibitem{brito2020}
I.~V.~S. Brito, L.~Sampaio, and L.~Zhang, ``Ndncomm 2020 poster: Towards a
  distance vector routing protocol for named data networking,'' \emph{Named
  Data Networking Community Meeting 2020}, 2020.

\bibitem{ndnwiki}
{NDN Project}, ``Namespace-based scope control, localhop scope,''
  \url{https://redmine.named-data.net/projects/nfd/wiki/ScopeControl},
  accessed: 2021-02-10.

\bibitem{perkins1994}
C.~E. Perkins and P.~Bhagwat, ``{Highly dynamic destination-sequenced
  distance-vector routing (DSDV) for mobile computers},'' \emph{ACM SIGCOMM
  computer communication review}, vol.~24, no.~4, pp. 234--244, 1994.

\bibitem{mohapatra2004}
P.~Mohapatra and S.~Krishnamurthy, \emph{AD HOC NETWORKS: technologies and
  protocols}.\hskip 1em plus 0.5em minus 0.4em\relax Springer Science \&
  Business Media, 2004.

\bibitem{li2019distributed}
T.~Li, Z.~Kong, S.~Mastorakis, and L.~Zhang, ``Distributed dataset
  synchronization in disruptive networks,'' in \emph{2019 IEEE 16th
  International Conference on Mobile Ad Hoc and Sensor Systems (MASS)}.\hskip
  1em plus 0.5em minus 0.4em\relax IEEE, 2019, pp. 428--437.

\bibitem{masf2020}
M.~Chowdhury, A.~Lane, and L.~Wang, ``Ndncomm 2020 presentation: m-asf - an
  adaptive srtt-based forwarding strategy for mobile environments,''
  \emph{Named Data Networking Community Meeting 2020}, 2020.

\end{thebibliography}

\end{document}